\documentclass[onecolumn,aps,floatfix,pra,superscriptaddress,nofootinbib]{revtex4}
\usepackage{color}
\usepackage{graphicx}
\usepackage{bm}
\usepackage{amsmath}
\usepackage{amsbsy}
\usepackage{hyperref}
\usepackage{enumerate}
\usepackage{upgreek}
\usepackage[subnum]{cases}
\usepackage{dsfont}

\newcommand{\tr}{ \text{tr} }
\newcommand{\ti}{ \tilde }
\newcommand{\ep}{ \epsilon }
\newcommand{\pa}{ \partial }
\newcommand{\hb}{ \hbar }
\newcommand{\si}{ \sigma }

\newcommand{\ga}{ \gamma }
\newcommand{\la}{ \langle }
\newcommand{\ra}{ \rangle }
\newcommand{\del}{ \delta }

\newcommand{\al}{ \alpha }
\newcommand{\be}{ \beta }

\newcommand{\rel}{ \text{rel.} }
\newcommand{\tho}{ \text{th} }
\newcommand{\epr}{ \text{EPR} }
\newcommand{\sech}{ \text{sech} }

\newcommand{\trho}{ \tilde{\rho} }

\newcommand{\Sch}{\text{Sch}}

\begin{document}

\title{ Dynamics of Quantum Correlations within the double Caldeira-Leggett formalism }

\author{S. V. Mousavi}
\email{vmousavi@qom.ac.ir}
\affiliation{Department of Physics, University of Qom, Ghadir Blvd., Qom 371614-6611, Iran}
\begin{abstract}

This study investigates the effects of decoherence and squeezing on the dynamics of various kinds of quantum features--local quantum coherence, local entropy, EPR correlations, and entanglement--in the high-temperature limit of the double Caldeira-Leggett model, focusing on initially squeezed states. We compare two scenarios: (1) particles interacting with distinct environments and (2) particles coupled to a common environment. Our analysis reveals that common environments better preserve local coherence over time, whereas distinct environments accelerate decoherence. Temperature enhances decoherence and suppresses coherence revivals, while squeezing affects transient dynamics but not long-term coherence saturation. Local entropy increases with temperature and squeezing, though their underlying physical mechanisms differ. EPR correlations degrade due to environmental interactions, with squeezing initially enhancing them but failing to prevent their eventual loss. Entanglement exhibits distinct behaviors: in separate environments, it undergoes sudden death, whereas in common environments, it experiences a dark period whose duration shortens with stronger squeezing. These findings provide a comprehensive understanding of how decoherence and squeezing influence quantum correlations in open quantum systems.

\end{abstract}

\maketitle




\section{Introduction}

Research in quantum information processing has taken two distinct paths: one focusing on discrete variables (qubits) and the other on high-dimensional, continuous-variable (CV) states such as coherent and squeezed states. These two approaches have begun to converge, leading to the development of potentially more powerful hybrid protocols \cite{Anetal-NP-2015}.
As a vital area of study in quantum information science and fundamental physics, CV quantum systems offer a robust platform for exploring quantum phenomena, including entanglement, quantum coherence, EPR correlations and entropy. The ability to manipulate and analyze these correlations has practical implications for quantum technologies such as communication, sensing, detection, imaging and computation.
These systems can be exemplified by quantized modes of bosonic systems, which encompass various degrees of freedom in the electromagnetic field, vibrational modes in solids, atomic ensembles, nuclear spins within a quantum dot, Josephson junctions, and Bose-Einstein condensates \cite{WePiGaCeRaShLl-RMP-2012}.

One of the essential tools for generating and manipulating quantum correlations in CV systems is the use of squeezed states. Squeezed states \cite{Lv-SL-2015} exhibit reduced quantum uncertainty in one quadrature at the expense of increased uncertainty in the conjugate quadrature, making them valuable for precision measurements and quantum information protocols. Gaussian states, a broader class of states characterized by Gaussian-shaped Wigner functions, provide a mathematically tractable framework for analyzing quantum correlations and are extensively utilized in experimental and theoretical studies \citep{AdIl-JPA-2007, LoSh-QI-2016}.

Various measures of non-classical correlations in CV systems have been explored in the literature. The primary measure is entanglement, which has been characterized and quantified in several ways. Consequently, distinguishing between separable and inseparable states is a fundamental task in quantum information science. Simon \cite{Si-PRL-2000} proposed a separability criterion for CV quantum systems that utilizes the geometric properties of the Peres-Horodecki partial transpose criterion. Furthermore, a criterion for inseparability in these systems has been introduced, based on the total variance of a pair of Einstein-Podolsky-Rosen (EPR) states \cite{DuGiCiZo-PRL-2000}. Among the most commonly used measures of entanglement in CV systems is entanglement negativity, which is defined as the smallest eigenvalue of the partially transposed covariance matrix \cite{Is-JRLR-2010}. This measure serves as a reliable indicator of the strength of quantum entanglement and is essential for characterizing the non-classicality of two-particle states.

Beyond entanglement, the negativity of the Wigner distribution function has emerged as an indicator of non-classical correlations. A negative Wigner function unambiguously signifies the departure from classical behavior and highlights the presence of genuine quantum features, such as coherence and superposition \cite{KeZy-JPB-2004, Hu-RMP-1974, Za-PRL-2024}.
The coherence and purity of reduced states offer further insights into the nature of quantum correlations. Quantum coherence, which reflects the ability of a system to exhibit superposition, is fundamental to quantum information processing \cite{BaCrPl-PRL-2014}. 
The relationship between local quantum coherence, quantified by skew information, and local quantum uncertainty as a measure of quantum correlations has been analyzed \cite{Gi-PRL-2014}. We will show that local $\ell_1$-norm quantum coherence is proportional to the coherence length, which corresponds to the width of the density matrix in the off-diagonal direction.
The purity of a reduced state, defined as a measure of its deviation from a completely mixed state, provides additional information on the extent of quantum coherence and entanglement within the system. Specifically, a practical experimental approach for estimating continuous variable entanglement has been proposed, relying on measurements of both global and marginal purities \cite{AdSeIl-PRL-2004, AdSeIl-PRA-2004}.
A key manifestation of quantum correlations in CV systems is the presence of EPR correlations \cite{EiPoRo-PR-1935}, which signify strong correlations between the quadratures of entangled particles \cite{Re-PRA-1989, GiWoKrWeCi-PRL-2003, RiEs-PRA-2004, MaGiViTo-PRL-2002, RuGoToWa-PRL-2013}. These correlations are often analyzed using covariance matrices, which succinctly capture the second-order statistical properties of a system's quadratures. 

Decoherence, a major challenge in maintaining quantum correlations, arises from interactions with the environment. Understanding decoherence dynamics is critical for preserving quantum features in practical applications. 
The Caldeira-Leggett (CL) formalism \cite{Ca-book-2014} is a widely used framework in the study of quantum dissipation and decoherence, particularly in systems interacting with their environments \cite{MoMi-EPJP-2020, MoMi-EPJP-2020_2, MoMi-Entropy-2021, MoMi-EPJP-2022}. It models a quantum system (like a particle) coupled to a large heat bath composed of harmonic oscillators. This approach has been instrumental in exploring how quantum coherence is lost due to environmental interactions, especially at high temperatures.
The double CL formalism \cite{CaMoPo-PA-2010} is an extension of the original CL model to the case of two quantum systems. In this framework, one is interested in understanding how the interaction of two particles with their environments leads to dissipation and decoherence, and how the nature of the environment---whether it is distinct (each system has its own independent bath) or common (both systems share the same bath)---affects their dynamics and mutual correlations.
%
In the distinct environment scenario, the two systems lose quantum coherence independently. In contrast, the common bath can induce correlations between the systems, sometimes even leading to the preservation or amplifying of entanglement despite the presence of noise.

In this research, we investigate the dynamics of quantum correlations in two-particle CV systems where the system is initially described by the squeezed state. The effect of decoherence on various quantum features will be studied in the framework of the double CL equation for both the distinct and common environments scenarios. Evolution of EPR correlations, log negativity, local quantum coherence and local purity will be examined for different values of squeezing parameter and also temperature of the environment. Through this comprehensive exploration, we aim to deepen the understanding of quantum correlations and their resilience in realistic physical environments. It is remarkable that our work is different than \cite{Bhetal-PRD-2023} where the authors have modeled the environment by a single one-dimensional free bosonic field. Explicitly, the system has been utilized by the researchers is a harmonic oscillator that is coupled to a free field. Notably, entanglement dynamics in continuous variable systems under decoherence has also been explored in other models, including those involving non-Markovian environments \cite{PB-JPA-2004, LiGo-PRA-2007, PaRo-PRL-2008}.

The remainder of the paper is structured as follows. Section \ref{sec: CL} introduces the Caldeira-Leggett formalism and its solution using the method of characteristics for a squeezed state. Section \ref{sec: decoh-eff} explores the effects of decoherence on various quantities and correlations, including local quantum coherence, local linear entropy, EPR correlations, and entanglement, providing analytical results whenever possible. In Section \ref{sec: res-discus}, we present and discuss our results. Finally, Section \ref{sec: sum-con} summarizes our findings and conclusions.

\section{The Caldeira-Leggett Formalism. Solution for the squeezed state} \label{sec: CL}

The Caldeira-Leggett formalism models a quantum system coupled to a large heat bath composed of harmonic oscillators. This approach has been used in exploring how quantum coherence is lost due to environmental interactions, especially at high temperatures. The bath degrees of freedom are traced out to derive an effective description of the system. This yields a dissipative term and introduces noise into the system dynamics. In this limit the master equation describing the system reads \cite{Ca-book-2014}
\begin{eqnarray} \label{eq: CL-oneparticle}
\frac{\pa \rho}{\pa t} &=& \frac{1}{i \hb} [H_0, \rho] + \frac{\ga}{i\hb} [x, \{ p,\rho \}] - \frac{D}{\hb^2} [x, [x,\rho]];
\end{eqnarray}
$H_0$ being the Hamiltonian of the system in consideration, $\ga$ is the dissipation constant and $D$ is the diffusion coefficient proportional to temperature,
\begin{eqnarray} \label{eq: diff-coeff}
 D &=& 2 m \ga k_B T 
\end{eqnarray}

In generalization of this scheme to a two-particle system, two situations may arise. In the first situation each particle interacts with its own environment i.e., the case of distinct environments. Here, the double CL equation for free non-interacting particles simply reads \cite{CaMoPo-PA-2010, MoMi-EPJP-2020, MoMi-EPJP-2022}
\begin{eqnarray} \label{eq: CL-distinct-1}
\frac{\pa}{\pa t}\rho(x_1, y_1; x_2, y_2; t) &=& \sum_{n=1}^2 \bigg[ 
+ i \frac{\hb}{2 m} \left( \frac{\pa^2}{\pa x_n^2} - \frac{\pa^2}{\pa y_n^2} \right) 
- \ga_n(x_n-y_n) \left( \frac{\pa}{\pa x_n} - \frac{\pa}{\pa y_n} \right) 
\nonumber \\
&~& \qquad - \frac{D_n}{\hb^2} (x_n-y_n)^2
\bigg]\rho(x_1, y_1; x_2, y_2; t)
\end{eqnarray}
where $\ga_n$ and $T_n$ imply respectively to the damping rate due to and the temperature of the $n^{\tho}$ environment; and $ D_n = 2 m \ga_n k_B T_n $.
The second scenario, which is more involved, arises when both particles interact with a common environment. 
The total Hamiltonian of the two-particle system coupled to a common environment is given by $ H = H_{S_1}+H_{S_2}+H_B+H_{S_1B}+H_{S_2B} $ where $H_{S_j}$ represents the Hamiltonian of the $j^{\text{th}}$ particle, $H_B$ is the Hamiltonian of the common bath, and $ H_{S_jB} $ describes the coupling between particle $j$ and the bath, given by $ H_{S_j B} = - x_j \sum_k c_{k,j} q_k $.  
After tracing out the bath degrees of freedom, the influence on the two particles is not merely the sum of independent environmental effects. Instead, cross terms emerge due to the shared environmental degrees of freedom coupling to both $x_1$ and $x_2$. These terms generate correlated noise and dissipation kernels, effectively inducing an interaction between the two particles, even in the absence of a direct coupling between them \cite{CaMoPo-PA-2010}.
%
As such, the double CL equation  takes the form 
\begin{eqnarray} \label{eq: CL-common-1}
\frac{\pa}{\pa t}\rho(x_1, y_1; x_2, y_2; t) &=& \sum_{n=1}^2 \bigg[ 
+ i \frac{\hb}{2 m} \left( \pa_{x_n}^2 - \pa_{y_n}^2 \right) 
- \ga (x_n-y_n) \left( \pa_{x_n} - \pa_{y_n} \right) 
- \frac{D}{\hb^2} (x_n-y_n)^2
\bigg]\rho(x_1, y_1; x_2, y_2; t) 
\nonumber \\
&-&
\bigg[ \ga \sum_n \sum_{n' \neq n } (x_n - y_n) \left( \pa_{x_{n'}} - \pa_{y_{n'}} \right)
+ 2 \frac{D}{\hb^2} (x_1 - y_1)(x_2 - y_2) \bigg] \rho(x_1, y_1; x_2, y_2; t)
\end{eqnarray}
where $ D $ is given by \eqref{eq: diff-coeff} and $ \pa_z $ represents partial derivative with respect to $z$. Note that the second line of this equation implies the effective interaction between particles induced by the common environment.

Since our goal is to examine various quantum properties, including correlations and quantum coherence, in our CV system, it is convenient to work with dimensionless quantities. To achieve this, we introduce a characteristic length scale, $\si_0$, which may correspond, for example, to the width of the initial wavepacket. Using this reference, we define the following dimensionless variables:
\begin{numcases}~
x \to \frac{x}{\si_0} \\
p \to \frac{ \si_0 }{ \hb } p \\ 
t \to \frac{ \hb }{ m \si_0^2 } t \\
\ga \to \frac{ m \si_0^2 }{ \hb } \ga \\
T \to \frac{ m \si_0^2 }{ \hb^2 } k_B T \\
\rho \to \frac{\rho}{\si_0^2}
\end{numcases}
which correspond to the position, momentum, time, damping rate, temperature, and density matrix, respectively. Assuming identical masses for both particles and identical environmental properties, CL equation for distinct environments, Eq. \eqref{eq: CL-distinct-1}, takes the form
\begin{eqnarray} \label{eq: CL-distinct}
\pa_t \rho(x_1, y_1; x_2, y_2; t) &=& \sum_{n=1}^2 \bigg[ 
\frac{i}{2} \left( \pa_{x_n}^2 - \pa_{y_n}^2 \right) - \ga(x_n-y_n) \left( \pa_{x_n} - \pa_{y_n} \right) 
- D (x_n-y_n)^2 \bigg]\rho(x_1, y_1; x_2, y_2; t)
\end{eqnarray}
in the dimensionless formulation, where $ D = 2 \ga T $.
For the common environment case we have that
\begin{eqnarray} 
\pa_t \rho &=& 
 \sum_{n=1}^2 \bigg[ 
\frac{i}{2} \left( \pa_{x_n}^2 - \pa_{y_n}^2 \right) - \ga(x_n-y_n) \left( \pa_{x_n} - \pa_{y_n} \right) 
- D (x_n-y_n)^2
\bigg]\rho 
\nonumber \\
& - & \bigg[ \ga (x_1-y_1)( \pa_{x_2} - \pa_{y_2} ) + \ga (x_2-y_2)( \pa_{x_1} - \pa_{y_1} ) + 2 D (x_1-y_1) (x_2-y_2)
\bigg] \rho \label{eq: CL-common}
\end{eqnarray}
Note that the second line of \eqref{eq: CL-common} displays additional effects of the common environment in comparison to the case of distinct environments.
We aim to consider influences of these additional terms on different quantum correlations and quantum coherence.

To solve CL equations \eqref{eq: CL-distinct} and \eqref{eq: CL-common}, we first do a transformation from coordinates $ (x_1, y_1; x_2, y_2) $ to the relative and center of mass coordinates $ (r_1, R_1; r_2, R_2) $ where $ r_i = x_i - y_i $ and $ R_i = (x_i + y_i)/2 $. In this way we get
\begin{eqnarray} 
\pa_t \rho &=& \left[
i ( \pa^2_{r_1, R_1} + \pa^2_{r_2, R_2} ) - 2 \ga ( r_1 \pa_{r_1} + r_2 \pa_{r_2} ) - D (r_1^2+r_2^2)
\right] \rho (r_1, R_1; r_2, R_2; t) \label{eq: CL-dis-rR}
\\
\pa_t \rho &=& \left[
i ( \pa^2_{r_1, R_1} + \pa^2_{r_2, R_2} ) - 2 \ga(r_1+r_2)( \pa_{r_1} + \pa_{r_2} ) - D (r_1+r_2)^2
\right] \rho (r_1, R_1; r_2, R_2; t) \label{eq: CL-common-rR}
\end{eqnarray}
respectively for the distinct environments scenario and the common environment case.
Then, partial Fourier transforms are done from coordinates $ (R_1, R_2) $ to $ (Q_1, Q_2) $;
\begin{eqnarray} \label{eq: ft-common}
\rho(r_1, R_1; r_2, R_2; t) & \to & \trho(r_1, Q_1; r_2, Q_2; t)
= \frac{1}{\sqrt{2\pi}}  \frac{1}{\sqrt{2\pi}} \int_{-\infty}^{\infty} \int_{-\infty}^{\infty} dR_1 dR_2 ~ e^{ i ( Q_1 R_1 + Q_2 R_2 ) } \rho(r_1, R_1; r_2, R_2; t)
\end{eqnarray}
Inserting the inverse partial Fourier transform into the double CL equations \eqref{eq: CL-dis-rR} and \eqref{eq: CL-common-rR} yields
\begin{eqnarray} 
\pa_t \trho &=&
- \left( (2 \ga r_1 - Q_1 ) \pa_{r_1} + ( 2\ga r_2 - Q_2 ) \pa_{r_2} + D (r_1^2 + r_2^2)  \right) \trho 
\label{eq: CL-dis-1_inverse}
\\
\pa_t \trho &=&
- \left( \left( 2 \ga (r_1+r_2) - Q_1 \right) \pa_{r_1} + \left( 2\ga(r_1+r_2) - Q_2 \right) \pa_{r_2}  + D (r_1+r_2)^2  \right) \trho 
\label{eq: CL-common-1_inverse}
\end{eqnarray}
respectively for the distinct and common environments.
These equations can be solved by the method of characteristics. By defining the curves
\begin{numcases}~
r_n = r_n(s) , \quad n= 1, 2 \\
t = t(s) ,
\end{numcases}
It follows that the system of coupled ordinary differential equations
\begin{numcases}~
\frac{d r_1}{ds} = 2 \ga r_1 - Q_1 ,  \label{eq: r1_char-dis}\\
\frac{d r_2}{ds} = 2 \ga r_2 - Q_2 , \label{eq: r2_char-dis} \\
\frac{d \trho}{ds} = - D ( r_1^2 +r_2^2) \trho , \label{eq: rhotilde_char-dis}\\
\frac{d t}{ds} = 1  \label{eq: s-t-dis},
\end{numcases}
is equivalent to the partial differential equation \eqref{eq: CL-dis-1_inverse}. Similarly, the system
\begin{numcases}~
\frac{d r_1}{ds} = 2 \ga ( r_1 + r_2 ) - Q_1 ,  \label{eq: r1_char-com}\\
\frac{d r_2}{ds} = 2 \ga ( r_1 + r_2 ) - Q_2 , \label{eq: r2_char-com} \\
\frac{d \trho}{ds} = - D (r_1+r_2)^2 \trho , \label{eq: rhotilde_char-com}\\
\frac{d t}{ds} = 1  \label{eq: s-t-com},
\end{numcases}
corresponds to the partial differential equation \eqref{eq: CL-common-1_inverse}.
First, coupled differential equations \eqref{eq: r1_char-dis} and \eqref{eq: r2_char-dis} are solved with the initial conditions $ r_1(0) = \xi_1 $ and $ r_2(0) = \xi_2 $. This yields expressions for $r_1$ and $r_2$ as functions of $ \xi_1, \xi_2, Q_1, Q_2 $ and $s$ i.e., $ r_1 = f_1(\xi_1, \xi_2, Q_1, Q_2, s) $ and $ r_2 = f_2(\xi_1, \xi_2, Q_1, Q_2, s) $. 
These solutions are then substituted into \eqref{eq: rhotilde_char-dis} to solve for $ \trho(s) $, given the initial condition $ \trho(r_1, Q_1; r_2, Q_2;0) = \trho_0(\xi_1, Q_1; \xi_2, Q_2) $.
%
%
Next, the equations for $ r_1 $ and $ r_2 $, are inverted to express $\xi_1$ and $\xi_2$ in terms of $ r_1 $, $ r_2 $, $Q_1$, $Q_2$ and $s$. These expressions are then substituted back into the solution for \eqref{eq: rhotilde_char-dis}, yielding $ \trho = \trho(r_1, Q_1; r_2, Q_2; s) $. Finally, using Eq. \eqref{eq: s-t-dis}, the parameter $ s$ is replaced by $t$, and partial inverse Fourier transforms are applied to obtain $ \rho(r_1, R_1; r_2, R_2; t) $. The same procedure is followed for the case of a common environment.

As we need first moments of both position and momentum operators to evaluate the entanglement content of our system, we consider now the Wigner distribution function $ W(x_1, p_1; x_2, p_2; t) $ which is defined as the partial Fourier transform of the the density matrix with respect to the relative coordinates $r_1$ and $r_2$ \cite{MoMi-Sym-2023},
\begin{eqnarray} \label{eq: wig-dis}
W(R_1, u_1; R_2, u_2; t) &=& \frac{1}{(2\pi)^2} \int dr_1 \int dr_2 e^{ - i (u_1 r_1 + u_2 r_2) } \rho(r_1, R_1; r_2, R_2; t).
\end{eqnarray}
Using this equation in the CL equation \eqref{eq: CL-common-rR}, yields the equation of motion
\begin{eqnarray} \label{eq: wig-eq}
\frac{\pa}{\pa t} W  &=& 
 \left[ \sum_{n=1}^2
 \left( - u_n \frac{\pa}{\pa R_n} + 2 \ga \frac{\pa}{\pa u_n} u_n + D \frac{\pa^2}{\pa u_n^2} \right)
+ 2 \ga \left( \frac{\pa}{\pa u_1} + \frac{\pa}{\pa u_2} \right) + 2 D \frac{\pa^2}{\pa u_1 \pa u_2} \right] W 
\end{eqnarray}
governing the Wigner distribution function \eqref{eq: wig-dis} for the case of the common environment. For the distinct environments the last two terms are absent.

\subsection{Solution for the squeezed state. Covariance matrix}

The initial state is taken the squeezed state whose representations in the position and momentum spaces reads respectively \cite{Lv-SL-2015} \footnote{Some authors have used 
$
\psi(x_1, x_2) = \sqrt{ \frac{2}{\pi} } \exp \left[ - e^{-2s} \frac{(x_1+x_2)^2}{2}  - e^{2s} \frac{(x_1-x_2)^2}{2} \right],
\phi(p_1, p_2) = \sqrt{ \frac{2}{\pi} } \exp \left[ - e^{-2s} \frac{(p_1-p_2)^2}{2}  - e^{2s} \frac{(p_1+p_2)^2}{2} \right], 
$
but as such $\phi(p_1, p_2)$ is not the Fourier transform of $\psi(x_1, x_2)$.
}
\begin{eqnarray}
\Psi_0(x_1, x_2) &=& \frac{1}{\sqrt{\pi} } \exp \left[ - e^{-2s} \frac{(x_1+x_2)^2}{4}  - e^{2s} \frac{(x_1-x_2)^2}{4} \right]
\label{eq: squ0_pos}
\\
\Phi_0(p_1, p_2) &=&  \frac{1}{\sqrt{\pi} } \exp \left[ - e^{-2s} \frac{(p_1-p_2)^2}{4}  - e^{2s} \frac{(p_1+p_2)^2}{4} \right]
\label{eq: squ0_mom}
\end{eqnarray}
$s$ being the squeezing parameter. In the limit of infinite squeezing, $ s \to \infty $, these functions asymptotically approach $ C \del(x_1-x_2) $ and $ C \del(p_1+p_2) $, respectively. These expressions correspond to the idealized correlations proposed by Einstein, Podolsky, and Rosen in their seminal paper \cite{EiPoRo-PR-1935}, where they argued for the incompleteness of the quantum description of physical reality.
In fact since Dirac delta states are unnormalizable and unphysical, researchers have proposed regularized versions \eqref{eq: squ0_pos} and \eqref{eq: squ0_mom}.
Furthermore, in the non-squeezed case ($ s=0 $), the state \eqref{eq: squ0_pos} is separable, as it corresponds to the product of two motionless Gaussian wave packets of equal width, both centered at the origin.
 
We have solved double CL equations \eqref{eq: CL-dis-rR} and \eqref{eq: CL-common-rR} analytically when the initial state is taken the squeezed state \eqref{eq: squ0_pos} but since solutions are lengthy we avoid to bring them here except for the Schr\"odinger case which the length is reasonable. Furthermore, the corresponding Wigner distribution functions are in hand. But, we only give the reduced or local state and the elements of the covariance matrix. It is important to note that for distinct environments, an initially separable state remains separable at all times.

\subsubsection{Solution of the Schr\"odinger equation}

The solution of the coupled Caldeira-Leggett equations \eqref{eq: CL-dis-rR} and \eqref{eq: CL-common-rR} in the limits of zero dissipation and zero temperature simplifies to the solution of the Schr\"odinger equation, which reads
\begin{eqnarray} \label{eq: Sch-sol}
\rho(r_1, R_1; r_2, R_2; t) &=& \frac{ 1 }{ \sqrt{ 1 + 2 \cosh(2s) t^2 + t^4 } }
\exp \left[ \frac{ \al(r_1; r_2, R_2; t) }{ \be(t) } \right],
\end{eqnarray}
where
\begin{eqnarray}
\be(t) &=& 8 \left(e^{4 s}+t^2\right) \left(e^{4 s} t^2+1\right) \\
\al(r_1; r_2, R_2; t) &=& 8 i e^{4 s} t^3 (r_1 R_1+r_2 R_2)-e^{2 s} \left(t^2 \left((r_1-r_2)^2+4 (R_1-R_2)^2\right)+(r_1+r_2)^2+4 (R_1+R_2)^2\right)
\nonumber \\
&-& e^{6 s} \left(t^2 \left((r_1+r_2)^2+4 (R_1+R_2)^2\right)+(r_1-r_2)^2+4 (R_1-R_2)^2\right)
\nonumber \\
&+& 4 i e^{8 s} t (r_1-r_2) (R_1-R_2)+4 i t (r_1+r_2) (R_1+R_2) .
\end{eqnarray}

\subsubsection{Reduced state}

Solution of the double CL equation with the initial squeezed state for both types of environments is symmetric under the exchange of particles, $ \rho(r_1, R_1; r_2, R_2; t) = \rho(r_2, R_2; r_1, R_1; t) $. Thus, the form of the reduced state is the same for both particles. In the case of distinct environments one obtains
\begin{eqnarray} 
\rho_A(r, R, t) &=& \int dr_2 \int dR_2 ~ \rho(r, R, r_2, R_2, t) \label{eq: loc-state} \\
&=& N_d(t) ~ \exp \left[ \frac{ d_{02}(t) R^2 + d_{11}(t) R r + d_{20}(t) r^2 }{ d_d(t) } \right] \label{eq: loc-state-dis}
\end{eqnarray}
where
\begin{equation}
N_d(t) = \frac{ 2\sqrt{2} \ga e^{2 \gamma  t} e^{-s} } 
{ \parbox{4.2in} {$ \big[ \pi \left(\left(4 e^{4 t \gamma } \gamma ^2-2 e^{2 t \gamma }+e^{4 t \gamma } + 1 \right) + \left( 4 e^{4 t \gamma } \gamma ^2-2 e^{2 t \gamma }+e^{4 t \gamma }+1\right) e^{-4s} \right) + $  \\ 
\hspace*{2.1cm} $
\left(\left(4 e^{4 t \gamma } (4 t \gamma -3)+16 e^{2 t \gamma }-4\right) e^{-2s} \right) T \big]^{1/2} $} }  
\end{equation}
and
\begin{eqnarray}
d_{02}(t) &=& -16  \gamma ^2 e^{4 \gamma  t} e^{-4s} ,\\
d_{11}(t) &=& 4 i \gamma  \left(e^{2 \gamma  t}-1 \right) e^{-2s} + 4 i \gamma  \left(e^{2 \gamma  t}-1 \right) e^{-6s}
+ 16 i \gamma  \left(e^{2 \gamma  t}-1\right)^2 e^{-4s} T , \\
d_{20}(t) &=& - \ga^2( 1 + 2 e^{-4s} + e^{-8s} ) + \bigg\{
\left(4 \gamma ^2-4 \gamma ^2 e^{4 \gamma  t}-4 \gamma  t+4 e^{2 \gamma  t}-e^{4 \gamma  t}-3\right) e^{-2s}  
\nonumber \\
& & + 
\left(4 \gamma ^2-4 \gamma ^2 e^{4 \gamma  t}-4 \gamma  t+4 e^{2 \gamma  t}-e^{4 \gamma  t}-3 \right) e^{-6s}
\bigg \} T ,
\\
d_d(t) &=& 2 \left(4 e^{4 t \ga } \ga ^2-2 e^{2 t \ga }+e^{4 t \ga }+1\right) e^{-2s} +2 \left(4 e^{4 t \ga } \ga ^2-2 e^{2 t \ga }+e^{4 t \ga }+1\right) e^{-6s} \nonumber \\
& & + \left[ 2 \left(4 e^{4 t \ga } (4 t \ga -3)+16 e^{2 t \ga }-4\right) e^{-4s} \right] T .
\end{eqnarray}
In the case of common environments one has that
\begin{eqnarray} \label{eq: loc-state-com}
\rho_A(r, R, t) &=& N_c(t) ~ \exp \left[ \frac{ c_{02}(t) R^2 + c_{11}(t) R r + c_{20}(t) r^2 }{ d_c(t) } \right]
\end{eqnarray}
where
\begin{eqnarray}
N_c(t) = \frac{ 4 \sqrt{2} \ga e^{4 \gamma  t} e^{-s} }
{ [ \pi \left(16 e^{8 t \gamma } \left(t^2+1\right) \gamma ^2+\left(e^{8 t \gamma } \left(16 \gamma ^2+1\right)-2 e^{4 t \gamma }+1\right) e^{-4s} \right) + \left(\left(2 e^{8 t \gamma } (8 t \gamma -3)+8 e^{4 t \gamma }-2\right) e^{-2s} \right) T ]^{1/2} }
\end{eqnarray}
and
\begin{eqnarray}
c_{02}(t) &=& -256 \gamma ^2 e^{8 \gamma  t} e^{-4s} , \\
c_{11}(t) &=& 128 i \gamma ^2  t e^{8 \gamma  t}  e^{-2s} + 32 i \gamma  \left(e^{4 \gamma  t}-1\right) e^{-6s} + 64 i \gamma  \left(e^{4 \gamma  t}-1\right)^2 e^{-4s} T ,
\\
c_{20}(t) &=& -16 \ga^2 e^{8 \ga t} + \left[ -16 \gamma ^2-16 \gamma ^2 t^2+\left(-16 \gamma ^2-1\right) e^{8 \gamma  t}-8 \gamma  t + 2 e^{4 \gamma  t} (4 \gamma  t+1) -1 \right] e^{-4s} - 16 \ga^2 e^{-8s}
\nonumber \\
& + & \bigg \{ 
\left[ 2 \left(16 \gamma ^2+16 \gamma ^2 t^2 + 8 \gamma  t+1\right)+e^{8 \gamma  t} \left(6-32 \gamma ^2 \left(t^2+1\right)\right)+8 e^{4 \gamma  t} (-4 \gamma  t-1) \right] e^{-2s} 
\nonumber \\
& & + \left[ 2 \left(16 \gamma ^2-8 \gamma  t-3\right)+\left(-32 \gamma ^2 - 2 \right) e^{8 \gamma  t}+8 e^{4 \gamma  t}  \right] e^{-6s} 
\bigg \} T 
\nonumber \\
&+& \left[-16 \left(\left(-1+e^{4 t \gamma }\right) \left(2 e^{4 t \gamma } t \gamma +2 t \gamma -e^{4 t \gamma }+1\right)\right)  \right] e^{-4s} T^2 ,
\\
d_c(t) &=& 128 e^{8 t \gamma } \left(t^2+1\right) \gamma ^2 e^{-2s} +8 \left(e^{8 t \gamma } \left(16 \gamma ^2+1\right)-2 e^{4 t \gamma }+1\right) e^{-6s}
\nonumber \\  
&+&  \left[ 8 \left(2 e^{8 t \gamma } (8 t \gamma -3)+8 e^{4 t \gamma }-2\right) e^{-4s} \right] T .
\end{eqnarray}
As can be easily seen, the numerator of the fraction in the exponent of the exponential function contains corrections that are quadratic in temperature, specifically the last term of $ c_{20}(t) $. In contrast, for distinct environments, the corrections are only linear in temperature. This difference influences the $\ell_1$-norm coherence, or the coherence length, which will be discussed later.

\subsubsection{Covariance matrix}

The covariance matrix in CV quantum systems is a crucial mathematical object that describes the statistical correlations of position and momentum, and it plays a central role in understanding the quantum nature of the system. 
For a two-mode Gaussian state, the $4 \times 4$ symmetric covariance matrix $\si$ is constructed out of three $2 \times 2 $ symmetric matrix
\begin{eqnarray} \label{eq: sig-mat}
\si &=&
\begin{pmatrix}
A & C \\
C^T & B
\end{pmatrix}
\end{eqnarray}
where $A$ and $B$ represent the correlators of the phase-space variables for subsystems 1 and 2, respectively, while $C$ denotes the cross-correlators between subsystems $A$ and $B$, 
\begin{eqnarray} \label{eq: cor-mat_A}
A &=&
\begin{pmatrix}
\la x_1^2 \ra - \la x_1 \ra^2 & \la x_1 p_1 \ra - \la x_1 \ra \la p_1 \ra \\
\\
\la x_1 p_1 \ra - \la x_1 \ra \la p_1 \ra & \la p_1^2 \ra - \la p_1 \ra^2 &
\end{pmatrix}
\qquad
C =
\begin{pmatrix}
\la x_1 x_2 \ra - \la x_1 \ra \la x_2 \ra & & \la x_1 p_2 \ra - \la x_1 \ra \la p_2 \ra \\
\\
\la p_1 x_2 \ra - \la p_1 \ra \la x_2 \ra & & \la p_1 p_2 \ra - \la p_1 \ra \la p_2 \ra &
\end{pmatrix}
\end{eqnarray}
%
%
For both types of environments, it is observed that always $ \la x_i \ra = \la p_i \ra = 0 $ for $ i= 1, 2 $ where $i$ labels the particles. Thus, the time-dependent covariance matrix reads
\begin{eqnarray} \label{eq: covmat}
\si(t) &=&
\left(
\begin{array}{cccc}
\la x_1^2 \ra_t & \la x_1 p_1 \ra_t & \la x_1 x_2 \ra_t & \la x_1 p_2 \ra_t 
\\
\\
\la x_1 p_1 \ra_t  & \la p_1^2 \ra_t & \la p_1 x_2 \ra_t & \la p_1 p_2 \ra_t 
\\ 
\\
\la x_1 x_2 \ra_t & \la p_1 x_2 \ra_t &  \la x_2^2 \ra_t & \la x_2 p_2 \ra_t 
\\ 
\\
\la x_1 p_2 \ra_t & \la p_1 p_2 \ra_t & \la x_2 p_2 \ra_t  & \la p_2^2 \ra_t \\
\end{array}
\right)
\end{eqnarray}
where
\begin{eqnarray} \label{eq: evaluate_mom}
\la x_i p_j \ra &=& \tr \left( \frac{x_i p_j + p_j x_i}{2} \rho \right) = \int\int\int\int dx_1dx_2dp_1dp_2 ~ x_i p_j ~ W(x_1, x_2, p_1, p_2) ,
\end{eqnarray}
$W$ being the Wigner distribution function.

In the Schr\"odinger framework, where our two-particle system is isolated and unaffected by any environment, the covariance matrix is given by
\begin{eqnarray}
\si_{\Sch}(t) &=&
\left(
\begin{array}{cccc}
 \frac{ 1+t^2 }{2} \cosh(2 s) & \frac{t}{2} \cosh (2 s) & \frac{ 1- t^2 }{2} \sinh(2s) & -\frac{t}{2} \sinh(2s) 
 \\
\\
 \frac{t}{2} \cosh (2 s) & \frac{1}{2} \cosh (2 s) & - \frac{t}{2} \sinh(2s) & -\frac{1}{2} \sinh(2s)
\\ 
 \\
 \frac{ 1- t^2 }{2} \sinh(2s) & - \frac{t}{2} \sinh(2s) &  \frac{ 1+t^2 }{2} \cosh(2 s) & \frac{t}{2} \cosh(2 s) 
\\ 
 \\
 - \frac{t}{2} \sinh(2s) & - \frac{1}{2} \sinh(2s) & \frac{ t }{2} \cosh (2 s) & \frac{1}{2} \cosh(2 s) \\
\end{array}
\right)
\end{eqnarray}
For the case of distinct environments the elements of the covariance matrix are given by
\begin{eqnarray}
\la x_1^2 \ra &=& \frac{\left(e^{4 s}+1\right) \left(\left(4 \gamma ^2+1\right) e^{4 \gamma  t}-2 e^{2 \gamma  t}+1\right) e^{-2 (s+2 \gamma  t)}}{16 \gamma ^2}-\frac{ -4 \gamma  t+e^{-4 \gamma  t}-4 e^{-2 \gamma  t}+3 }{4 \gamma ^2} ~T \\
\la x_1 p_1 \ra &=& \frac{\cosh (2 s) e^{-3 \gamma  t} \sinh (\gamma  t)}{2 \gamma }+\frac{2 e^{-2 \gamma  t} \sinh ^2(\gamma  t)}{\gamma }~T \\
\la p_1^2 \ra &=& \frac{1}{4} \left(e^{4 s}+1\right) e^{-2 (s+2 \gamma  t)}+ \left(1-e^{-4 \gamma  t}\right) T \\
\la x_1 x_2 \ra &=& \frac{\left(e^{4 s}-1\right) e^{-2 (s+2 \gamma  t)} \left(\left(4 \gamma ^2-1\right) e^{4 \gamma  t}+2 e^{2 \gamma  t}-1\right) }{16 \gamma ^2} \\
\la x_1 p_2 \ra &=& -\frac{\sinh (2 s) e^{-3 \gamma  t} \sinh (\gamma  t)}{2\gamma } \\
\la p_1 p_2 \ra &=& - \frac{ \sinh (2 s) }{2} e^{-4 \gamma  t} 
\end{eqnarray}
These equations indicate that thermal fluctuations have no impact on the cross-correlators, while the remaining correlators are influenced solely by the temperature to the first power.
When both particles interact with a common environment one has that
\begin{eqnarray}
\la x_1^2 \ra &=& \frac{e^{-2 (s+4 \gamma  t)} \left(e^{8 \gamma  t} \left(16 \gamma ^2 \left(e^{4 s} \left(t^2+1\right)+1\right)+1\right)-2 e^{4 \gamma  t}+1\right)}{64 \gamma ^2}-\frac{\left(-8 \gamma  t+e^{-8 \gamma  t}-4 e^{-4 \gamma  t}+3\right)}{32 \gamma ^2}~T \\
\la x_1 p_1 \ra &=& \frac{e^{-2 (s+4 \gamma  t)} \left(4 \gamma  t e^{4 s+8 \gamma  t}+e^{4 \gamma  t}-1\right)}{16 \gamma }+\frac{ e^{-8 \gamma  t} \left(e^{4 \gamma  t}-1\right)^2}{8 \gamma }~T \\
\la p_1^2 \ra &=& \frac{1}{4} e^{-2 s} \left(e^{4 s}+e^{-8 \gamma  t}\right)+ \left(\frac{1 - e^{-8 \gamma  t} }{2} \right) T 
\end{eqnarray}
for the elements of the covariance matrix of the subsystems 1 and 2; and
\begin{eqnarray}
\la x_1 x_2 \ra &=& \frac{e^{-2 (s+4 \gamma  t)} \left(e^{8 \gamma  t} \left(1-16 \gamma ^2 \left(e^{4 s} \left(t^2-1\right)+1\right)\right)-2 e^{4 \gamma  t}+1\right)}{64 \gamma ^2}-\frac{ -8 \gamma  t+e^{-8 \gamma  t}-4 e^{-4 \gamma  t}+3 }{32 \gamma ^2} T \\
\la x_1 p_2 \ra &=& \frac{e^{-2 (s+4 \gamma  t)} \left(-4 \gamma  t e^{4 s+8 \gamma  t}+e^{4 \gamma  t}-1\right)}{16 \gamma } + \frac{ e^{-8 \gamma  t} \left(e^{4 \gamma  t}-1\right)^2}{8 \gamma }~T \\
\la p_1 p_2 \ra &=& \frac{1}{4} e^{-2 s} \left(e^{-8 \gamma  t}-e^{4 s}\right) + \left(\frac{1 - e^{-8 \gamma  t} }{2} \right) T 
\end{eqnarray}
for the cross correlators. In contrast to the case of distinct environments, thermal fluctuations in the common environment also influence the cross-correlators.

\section{Effect of decoherence on various quantities} \label{sec: decoh-eff}

In this section, we examine the dynamics of quantum coherence and the linear entropy of the reduced state, EPR correlations, and entanglement, which is quantified by logarithmic negativity, within the double CL framework for both distinct and common environment scenarios.

\subsection{$\ell_1$-norm of quantum coherence}

The $\ell_1$-norm coherence of a quantum state is defined as the sum of the magnitudes of the off-diagonal elements of the system's density matrix, i.e.,
\begin{eqnarray}
C_{\ell_1}(\hat{\rho}) &=& \sum_i \sum_{j \neq i} | \rho_{ij} | \label{eq: cl1norm_1}
\\
&=& \sum_i \sum_{j} | \rho_{ij} | - \sum_i |\rho_{ii}| . \label{eq: cl1norm_2}
\end{eqnarray}
%
%
Extending this relation to CV states is more complex and involves both physical and mathematical subtleties. From a physical perspective, it is important to consider the dimensions of the quantities involved. Coherence must be dimensionless, but when particles move in one dimension, the density matrix in the position representation has the dimension $ L^{-N} $, where $L$ is the length and $N$ is the number of particles. Therefore, to ensure meaningful results, one must work with dimensionless quantities.
From the mathematical point of view, integral of a well-defined function over a finite area, such as $ \int_a^b dx \int_c^d dy f(x, y) $, is zero along the line $ y = x $. However, when integrating over an infinite region, the situation becomes more complicated, and a detailed analysis is required. A limiting procedure should be applied to evaluate the integral along the line $ y = x $. For example, this integral can be computed over a strip of width $2 \ep$ around the line $ y = x $. In this way, for Eq. \eqref{eq: cl1norm_2} one has that  
\begin{eqnarray}
C_{\ell_1}(\hat{\rho}) &=& \int_{-\infty}^{\infty} dx \int_{-\infty}^{\infty} dy | \rho(x, y) | 
- \lim_{\ep \to 0} \int_{-\infty}^{\infty} dx \int_{x-\ep}^{x+\ep} dy | \rho(x, y) | .
\end{eqnarray}
But, since the density matrix is a bounded function, the second integral is zero and we simply have
\begin{eqnarray} \label{eq: l1-norm-CV}
C_{\ell_1}(\hat{\rho}) &=& \int_{-\infty}^{\infty} dx \int_{-\infty}^{\infty} dy | \rho(x, y) |.
\end{eqnarray}

We now give analytical relations for the $\ell_1$-norm coherence of the reduced states \eqref{eq: loc-state-dis} and \eqref{eq: loc-state-com} respectively for the distinct and common environments. 
For the case of distinct environments one has that 
\begin{eqnarray}
C_{\ell_1, d}(\hat{\rho}_A(t)) &=& \sqrt{ \frac{f_d(t)}{g_d(t)} } \label{eq: cl1-rhoA-dis}
\end{eqnarray}
where
\begin{eqnarray}
f_d(t) &=&  2 \pi  \left(e^{4 t \gamma } \left(4 \gamma ^2+1\right)-2 e^{2 t \gamma }+1\right) e^{-2s} +2 \pi  \left(e^{4 t \gamma } \left(4 \gamma ^2+1\right)-2 e^{2 t \gamma } + 1 \right)e^{-6s}
\nonumber \\
&+&
2 \pi  \left(4 e^{4 t \gamma } (4 t \gamma -3)+16 e^{2 t \gamma }-4\right) e^{-4s} ~ T ,
 \\
g_d(t) &=& 
\gamma^2 [ 1 + 2 e^{-4s} + e^{-8s}  ] + \big\{
[ -4 \gamma ^2+4 \gamma ^2 e^{4 \gamma  t}+4 \gamma  t-4 e^{2 \gamma  t}+e^{4 \gamma  t}+3 ] e^{-2s}
\nonumber \\
&+&
[ -4 \gamma ^2+4 \gamma ^2 e^{4 \gamma  t}+4 \gamma  t-4 e^{2 \gamma  t}+e^{4 \gamma  t}+3 ] e^{-6s}
\big\} T
+ [ 16 \gamma  t \left(e^{4 \gamma  t}-1\right)-16 \left(e^{2 \gamma  t}-1\right)^2 ] e^{-4s} T^2 ,
\end{eqnarray}
sub-index $d$ referring to distinct environments. 
It is worth-mentioning here to comment on the coherence length. This quantity, is the width of the density matrix in the off-diagonal direction, i.e., $ y = -x $ or $ r = 2x $. 
As such, the coherence length is defined as $ L(t) = 1 / \sqrt{- 8 A(t)} $ where $A(t)$ is the coefficient of $r^2$ in the exponent of the density matrix. One can see that 
\begin{eqnarray} \label{eq: cohlen}
 L(t) &=& 2 \sqrt{2\pi} C_{\ell_1}(\hat{\rho}_A(t)) .
\end{eqnarray}
The stationary value of the coherence \eqref{eq: cl1-rhoA-dis} reads
\begin{eqnarray} \label{eq: cl1-rhoA-dis_st}
C_{\ell_1, d}(\hat{\rho}_A(t)) \bigg|_{t \to \infty} &=& \sqrt{ \frac{2\pi}{T} } 
\end{eqnarray}
being independent of the squeezing parameter $s$. 
In the limit $ \ga \to 0, T \to 0 $ i.e., in the Schr\"odinger framework we obtain
\begin{eqnarray} \label{eq: cl1-rhoA-Sch}
C_{\ell_1, \Sch}(\hat{\rho}_A(t)) &=& 2 \sqrt{ \frac{\pi}{ \cosh(2s) } }~ \sqrt{1+t^2 }
\end{eqnarray}
for the $\ell_1$-norm coherence of the reduced state $\rho_A$, which can also be derived directly. In the Schr\"odinger framework, for the special case $s=0$, we have $ \rho_A(x, y, t) = \psi(x, t) \psi^*(y, t) $, where, for the initial state given by Eq. \eqref{eq: squ0_pos} with $s=0$, $ \psi(x, t) $ is a Gaussian wavepacket with an initial width $ 1 / \sqrt{2} $, and a time-dependent width $\si_t = \sqrt{ (1+t^2)/2 }$. Therefore, for the local coherence, we obtain
\begin{eqnarray} \label{eq: cl1-sch-s=0}
C_{\ell_1, \Sch}(\hat{\rho}_A(t)) \bigg |_{s=0} &=& \int dx \int dy | \psi(x, t) \psi^*(y, t) | = \left( \int dx | \psi(x, t)| \right)^2
= 2 \sqrt{2\pi} ~ \si_t.
\end{eqnarray}
Equation \eqref{eq: cl1-rhoA-Sch} shows that local coherence in the Schr\"odinger framework always increases with time, which is expected since the coherence length, i.e., the width of the density matrix in the off-diagonal direction, is an increasing function of time. In this case, for Gaussian wave packets, as seen in equations \eqref{eq: cohlen} and \eqref{eq: cl1-sch-s=0}, the coherence length is simply the ensemble width $ \si_t $.

In the case of common environments we have
\begin{eqnarray}
C_{\ell_1, c}(\hat{\rho}_A(t)) &=& \sqrt{ \frac{f_c(t)}{g_c(t)} } \label{eq: cl1-rhoA-com}
\end{eqnarray}
where
\begin{equation}
f_c(t) = \left(128 e^{8 t \gamma } \pi  \left(t^2+1\right) \gamma ^2 e^{-2 s} +8 \pi  \left(e^{8 t \gamma } \left(16 \gamma ^2+1\right)-2 e^{4 t \gamma }+1\right) e^{-6 s} \right)+ 8 \pi  \left(2 e^{8 t \gamma } (8 t \gamma -3)+8 e^{4 t \gamma }-2 \right) e^{-4 s}   T 
\end{equation}
and
\begin{eqnarray}
g_c(t) &=& 
16 \ga^2 e^{8\ga t} + [ 16 \gamma ^2+16 \gamma ^2 t^2+\left(16 \gamma ^2+1\right) e^{8 \gamma  t}+8 \gamma  t-2 e^{4 \gamma  t} (4 \gamma  t+1)+1 ] e^{-4 s} + 16 \ga^2 e^{-8 s} 
\nonumber \\
&+&
\bigg\{ [ -2 \left(16 \gamma ^2+16 \gamma ^2 t^2+8 \gamma  t+1\right)+e^{8 \gamma  t} \left(32 \gamma ^2 \left(t^2+1\right)-6\right)-8 e^{4 \gamma  t} (-4 \gamma  t-1) ] e^{-2 s}
\nonumber \\
&+&
[ -2 \left(16 \gamma ^2-8 \gamma  t-3\right)+\left(32 \gamma ^2+2\right) e^{8 \gamma  t}-8 e^{4 \gamma  t} ] e^{-6 s} \bigg\} T
\nonumber \\
&+&
16 \left(e^{4 \gamma  t}-1\right) \left(2 \gamma  t e^{4 \gamma  t}+2 \gamma  t-e^{4 \gamma  t}+1\right) e^{-4 s} ~ T^2 ;
\end{eqnarray}
sub-index $c$ referring to common environments.
The stationary value of the local coherence \eqref{eq: cl1-rhoA-com} reads
\begin{eqnarray} \label{eq: cl1-rhoA-com_st}
C_{\ell_1, c}(\hat{\rho}_A(t)) \bigg|_{t \to \infty} &=& 2 \sqrt{ \frac{\pi}{T} }
\end{eqnarray}
which is independent of the value of the squeezing parameter $s$. Furthermore, it is $ \sqrt{2} $ times larger than that of the distinct environments case. This suggests that a common environment preserves more coherence than distinct environments.
Here, also one has $ L(t) = 2 \sqrt{2\pi} C_{\ell_1}(\hat{\rho}_A(t)) $ for the coherence length which establishes a direct link between coherence and spatial coherence length in the system.

\subsection{ Purity of reduced states. Linear entropy}

Due to environmental interaction, any pure quantum state participating in a quantum information process evolves into a mixed state. As a result, an essential aspect of Quantum Information Theory is quantifying the mixedness of a quantum state. We quantify this degree with the linear entropy. The state of particle $A$ is obtained by tracing out the degrees of freedom of particle $B$ from the total density matrix, as described by Eq. \eqref{eq: loc-state}.
For the square of the reduced density matrix one has that 
\begin{eqnarray}
\rho_A^2(x, y, t) &=& \la x | \hat{\rho}_A^2(t) | y \ra =
\int_{-\infty}^{\infty} dz \rho_A(z, y, t) \rho_A(x, z, t).
\end{eqnarray}
A well-known quantity to characterize the degree of decoherence (or mixedness) that the subsystem $A$ experiences, due to subsystem $B$, is called purity. The purity of the reduced state \eqref{eq: loc-state} is given by
\begin{eqnarray} \label{eq: local_pur}
P(\hat{\rho}_A(t)) &=& \tr(\hat{\rho}_A^2(t)) =
\int_{-\infty}^{\infty} dx \int_{-\infty}^{\infty} dz ~ \rho_A(x, z, t) \rho_A^*(x, z, t),
\end{eqnarray}
and the corresponding linear entropy is defied as
\begin{eqnarray} \label{eq: local_linent}
S(\hat{\rho}_A(t)) &=& 1 - P(\hat{\rho}_A(t)).
\end{eqnarray}
The linear entropy varies in the range $ 0 \leq S(\hat{\rho}_A) \leq 1 - 1/D $ where $D$ is the dimension of the vector space being infinity for CV systems. Thus, the maximum value of the linear entropy is one for our CV system.

From \eqref{eq: local_pur} we obtain
\begin{equation} \label{eq: pure-dis}
P_d(\rho_A(t)) = \frac{ 2 \ga e^{2 \ga  t} e^{-2s} }
{ \parbox{5in} {$ \bigg[ 
\ga^2(1+2 e^{-4s} + e^{-8s} ) + 
\big\{ ( -4 \gamma ^2+4 \gamma ^2 e^{4 \gamma  t}+4 \gamma  t-4 e^{2 \gamma  t}+e^{4 \gamma  t}+3 ) (e^{-2s} + e^{-6s} )
\big\}T + $  \\ 
\hspace*{2.1cm} $
\left( 16 \gamma  t \left(e^{4 \gamma  t}-1\right)-16 \left(e^{2 \gamma  t}-1\right)^2 \right) e^{-4s} T^2
\bigg]^{1/2} $} } 
\end{equation}
for the purity of the local state \eqref{eq: loc-state-dis}, when particles interact with two distinct environments and
\begin{eqnarray}
P_c(\rho_A(t)) &=& \frac{ 8 \ga e^{4 \ga  t}  e^{-2s} }{\sqrt{b_c(t)} } \label{eq: pure-com}
\end{eqnarray}
for the case of common environment where
\begin{eqnarray}
b_c(t) &=& 16 \gamma ^2 e^{8 \gamma  t} + 
[ 8 \gamma  \left(2 \gamma +2 \gamma  t^2+t\right)+\left(16 \gamma ^2+1\right) e^{8 \gamma  t}-2 e^{4 \gamma  t} (4 \gamma  t+1)+1 ] e^{-4s} + 16 \ga^2 e^{-8s}
\nonumber \\
&+& 
\big\{
[ e^{8 \gamma  t} \left(32 \gamma ^2 \left(t^2+1\right)-6\right)+2 \left(-8 \gamma  \left(2 \gamma  \left(t^2+1\right)+t\right)-1\right)-8 e^{4 \gamma  t} (-4 \gamma  t-1) ] e^{-2s}
\nonumber \\
&+& [\left(32 \gamma ^2+2\right) e^{8 \gamma  t}+2 (8 \gamma  (t-2 \gamma )+3)-8 e^{4 \gamma  t}] e^{-6s}
\big\}T
+ 16 \left(e^{4 \gamma  t}-1\right) \left(2 \gamma  t e^{4 \gamma  t}+2 \gamma  t-e^{4 \gamma  t}+1\right) e^{-4s} T^2 .
\nonumber \\
\end{eqnarray}
In the limit $ \ga \to 0, T \to 0 $, Eqs. \eqref{eq: pure-dis} and \eqref{eq: pure-com} yields
\begin{eqnarray} \label{eq: pure-Sch}
P_{\Sch}(\rho_A(t)) &=& \sech(2s)
\end{eqnarray}
which is just the result in the Schr\"odinger framework. As expected, it is independent of time and has the maximum value, one, for the non-squeezed state, $s=0$.

In the short time limit, the local linear entropies, $ S(\rho_A) = 1 - \tr(\rho_A^2) $, are given by
\begin{eqnarray}
S_d(\rho_A(t)) & \approx & (1 - \sech(2s) ) + 2 \sech(2 s) ( 2 \sech(2 s)~T - 1 ) ~ \ga t + O(t^2) \label{eq: linent-short-dis} 
\end{eqnarray}
and
\begin{eqnarray}
S_c(\rho_A(t)) & \approx & (1 - \sech(2s) ) + 2 \sech(2 s) ( 1 - \tanh(2s) ) ( -1 + 2 e^{2s} T ) ~ \ga t + O(t^2) \label{eq: linent-short-com} 
\end{eqnarray}
for the distinct and common environments, respectively. From equation \eqref{eq: linent-short-dis}, it can be seen that for $ T < \cosh(2s)/2 $, the linear entropy for distinct environments initially decreases with time. However, for the common environment scenario, the time derivative of the local entropy in Eq. \eqref{eq: linent-short-com} is initially negative only when $ T < e^{-2s} /2 $, which does not hold in the high-temperature limit of the CL framework.

\subsection{EPR correlations}

Considering a two-particle system, Einstein, Podolsky and Rosen took the position-space wave function $ \Psi(x_1, x_2) = C \del(x_1-x_2+X) $ which is an eigenfunction of the relative position operator $ \hat{x}_1 - \hat{x}_2 $ with the eigenvalue $X$. The corresponding quantum state is \cite{LoSh-QI-2016}
\begin{eqnarray} \label{eq: epr-state}
|\epr \ra &=& \int \int dx_1 dx_2 |x_1, x_2\ra \la x_1, x_2 | \Psi \ra = C \int \int dx_1 dx_2 ~ \del(x_1-x_2+X) |x_1, x_2\ra
\end{eqnarray}
which is an eigenstate of the total momentum operator $ \hat{p}_1 + \hat{p}_2 $ wit eigenvalue zero;
\begin{eqnarray*}
(\hat{p}_1 + \hat{p}_2)|\epr \ra &=& \int \int dx_1 dx_2 |x_1, x_2\ra \la x_1, x_2 | (\hat{p}_1 + \hat{p}_2) | \Psi \ra
\\
&=& C \frac{\hb}{i} \int \int dx_1 dx_2 |x_1, x_2\ra (\pa_{x_1}+\pa_{x_2}) \frac{1}{2\pi \hb} \int dp ~ e^{ip(x_1-x_2+X)/\hb} = 0
\end{eqnarray*}
where in the second line we have used the representation of the Dirac delta function in terms of plane waves. This analysis show that the $| \epr \ra$ state \eqref{eq: epr-state} is a simultaneous eigenstate of the compatible global observables--relative position and total momentum--exhibiting perfect correlations in both positions and momenta.
Since this state is unnormalizable and unphysical, researchers have explored regularized versions that converge to it in the appropriate limit. The $| \epr \ra$ state can be thought of as the limiting case of the squeezed state \eqref{eq: squ0_pos} and \eqref{eq: squ0_mom}.
As such, the variances in relative position and total momentum have the same value as $ e^{-2s} $.

In \cite{DuGiCiZo-PRL-2000} and \cite{MaGiViTo-PRL-2002} sufficient conditions has been given for inseparability. It has been proved that for any separable quantum state $\rho$ the total variance of a pair of EPR-like operators $ \hat{r} = \hat{x}_1 - \hat{x}_2 $ and $ \hat{u} = \hat{p}_1 + \hat{p}_2 $, respectively the relative position and total momentum operators, satisfies the inequality \cite{DuGiCiZo-PRL-2000}
\begin{eqnarray}
\frac{ \del_{\hat{r}} + \del_{\hat{u}} }{2} \geq 1,
\end{eqnarray}
where $ \del_o = \la o^2 \ra - \la o \ra^2 $ is the uncertainty in measurement of the operator $o$. Note that this relation is meaningful only if the quantities are dimensionless.
Based on this proof, EPR correlation or EPR uncertainty has been defined as \cite{GiWoKrWeCi-PRL-2003, RiEs-PRA-2004}
\begin{eqnarray} \label{eq: EPR-uncer}
\xi &=& \frac{ \del_{x_1-x_2} + \del_{p_1+p_2} }{2} .
\end{eqnarray}
This quantity is a measure of nonlocal correlations; the existence of nonlocal correlations is implied by $ \xi < 1 $.
Note that for the ideal EPR state (the simultaneous eigenfunctions of relative coordinate and the total momentum) one has $ \xi = 0 $. 
This means that the more nonlocal a system is, the closer $\xi$ is to zero.
%

A very useful entanglement witness for continuous variables is the Mancini-Giovannetti-Vitali-Tombesi (MGVT) criterion \cite{MaGiViTo-PRL-2002, RuGoToWa-PRL-2013}. If the CV bipartite state $\rho$ is separable then 
\begin{eqnarray}
\del_{x_1-x_2} ~ \del_{p_1+p_2} \geq 1
\end{eqnarray}
This is a sufficient condition for inseparability as the authors have clearly written in their original paper. Then, the degree of entanglement has been defined as \cite{MaGiViTo-PRL-2002} 
\begin{eqnarray} \label{eq: eta}
\eta = \del_{x_1-x_2} ~ \del_{p_1+p_2}.
\end{eqnarray}
The condition $ \eta < 1 $ implies entanglement while $ \eta < 1/4 $ is the indication for EPR correlations \cite{ReDr-PRL-1988}.

For the common environment case we obtain
\begin{eqnarray}
\xi(t) &=& \frac{1}{2} \left(e^{2 s} t^2+e^{-2 (s+4 \gamma  t)}+e^{-2 s}\right) +  \left( 1 - e^{-8 \gamma  t}\right) T ,  \label{eq: xi-com} \\
\eta(t) &=& \left(e^{4 s} t^2+1\right) e^{-4 (s+2 \gamma  t)} + 2 \left(e^{2 s} t^2 + e^{-2 s} \right) \left( 1 - e^{-8 \gamma  t} \right) T , \label{eq: eta-com}
\end{eqnarray}
while for the case of distinct environments we get
\begin{eqnarray}
\xi(t) &=& \frac{e^{-2 (s+2 \gamma  t)} \left(4 \gamma ^2+\left(4 \gamma ^2+e^{4 s}\right) e^{4 \gamma  t}-2 e^{4 s+2 \gamma  t}+e^{4 s}\right)}{8 \gamma ^2}-\frac{ e^{-4 \gamma  t} \left(4 \gamma ^2-4 e^{2 \gamma  t}+1\right)-4 \gamma  (\gamma +t)+3 }{4 \gamma ^2} T ,
 \label{eq: xi-dis} 
 \\
\eta(t) &=& e^{-4 (s+\gamma  t)}+\frac{e^{-6 \gamma  t} \sinh ^2(\gamma  t)}{\gamma ^2}
\nonumber \\
&+& \frac{e^{-2 (s+4 \gamma  t)} \left(\left(4 \gamma ^2+e^{4 s}\right) e^{8 \gamma  t}+2 \left(e^{4 s}+2\right) e^{2 \gamma  t}-2 e^{4 s+6 \gamma  t}-e^{4 s}+e^{4 \gamma  t} (4 \gamma  (t-\gamma )-3)-1\right)}{2 \gamma ^2} ~ T
\nonumber \\
&+& \frac{e^{-8 \gamma  t} \left(e^{4 \gamma  t}-1\right) \left(e^{4 \gamma  t} (4 \gamma  t-3)+4 e^{2 \gamma  t}-1\right)}{\gamma ^2}~T^2 .
 \label{eq: eta-dis}
\end{eqnarray}
A comparison of Eq. \eqref{eq: eta-com} with Eq. \eqref{eq: eta-dis} reveals that in the common environment scenario, the EPR correlation 
$ \eta(t) $ is affected by thermal noise to the first power of temperature, whereas in the case of distinct environments, it is influenced to the second power of temperature.

In the Schr\"odinger framework i.e., in the absence of any environment one obtains
\begin{eqnarray}
\xi(t)\big|_{\Sch} &=&   e^{-2s} + \frac{1}{2} e^{2s} t^2    \label{eq: xi-Sch} \\
\eta(t)|_{\Sch} &=& e^{-4s} + t^2  \label{eq: eta-Sch}
\end{eqnarray}

\subsection{Entanglement negativity}

The Peres-Simon necessary and sufficient criterion for separability has been utilized \cite{Is-JRLR-2010} to investigate the entanglement dynamics of two-mode Gaussian states through the covariance matrix \eqref{eq: sig-mat}. The partial transposition of the bipartite Gaussian density matrix $\rho$ changes the covariance matrix $\si$ into a new matrix $\ti{\si}$ where the determinant of the cross correlator $C$ flips the sign. 
Entanglement negativity is given by the eigenvalues of symplectic eigenvalues $ \tilde{\nu} $ of the new covariance matrix $\ti{\si}$,
\begin{eqnarray} \label{eq: log-neg}
E_N &=& \max\{ 0, - \log_2(2 \ti{\nu}_-) \}
\end{eqnarray}
where $ \ti{\nu}_- $ is the smallest eigenvalue of the partially transposed covariance matrix $\ti{\si}$,
\begin{eqnarray} \label{eq: eigenvalues}
\ti{\nu}_{\mp} &=& \frac{1}{2} \left( \ti{\Delta} \mp \sqrt{ \ti{\Delta}^2 - 4 \det \si } \right). 
\end{eqnarray}
Here, $ \ti{\Delta} = \det A + \det B - 2 \det C $ is the symplectic invariant.
State $\rho$ is separable if and only if
\begin{eqnarray} \label{eq: sep-condition}
\ti{\nu}_- \geq \frac{1}{2}
\end{eqnarray}
and the logarithmic negativity \eqref{eq: log-neg} quantifies violation of this inequality. 

Equations for $ \ti{\nu}_- $ and $E_N$ are too lengthy to be presented here. Therefore, to illustrate the impact of thermal fluctuations on entanglement, we give solely the expressions for $ \ti{\Delta} $ and $ \det(\si) $.
In the case of common environment we have that
\begin{eqnarray}
\ti{\Delta} &=& \frac{ f_c(t) }{ 64 \gamma ^2 e^{-4s} } \\
\det(\si) &=& \frac{ g_c(t) }{ 128 \gamma ^2 e^{-2s} }
\end{eqnarray}
where
\begin{eqnarray}
f_c(t) &=& [ 16 \gamma ^2 + e^{-8 \gamma  t} \left(-4 \gamma  t+e^{4 \gamma  t}-1\right)^2 e^{-4s} +16 \gamma ^2 e^{-8 \gamma  t} e^{-8s} ]
\nonumber \\
& & + \big\{
[ e^{-8 \gamma  t} \left(-2 \left(16 \gamma ^2 t^2+8 \gamma  t+1\right)+e^{8 \gamma  t} \left(32 \gamma ^2 t^2-6\right)+8 e^{4 \gamma  t} (4 \gamma  t+1)\right) ] e^{-2s} + 32 \gamma ^2 e^{-8 \gamma  t} \left(e^{8 \gamma  t}-1\right) ~ e^{-6s}
\big\}T ,
\nonumber \\
\end{eqnarray}
and
\begin{eqnarray}
g_c(t) &=& 8 \gamma ^2 e^{-8 \gamma  t} e^{-2s} + [ 16 \gamma ^2 e^{-8 \gamma  t} \left(e^{8 \gamma  t}-1\right) +e^{-8 \gamma  t} \left(8 \gamma  t-4 e^{4 \gamma  t}+e^{8 \gamma  t}+3\right) e^{-4s}  ] T
\nonumber \\
& & +~ 8 e^{-8 \gamma  t} \left(e^{4 \gamma  t}-1\right) \left(2 \gamma  t+e^{4 \gamma  t} (2 \gamma  t-1)+1\right) e^{-2s}~ T^2.
\end{eqnarray}

\section{Results and discussions} \label{sec: res-discus}

\begin{figure} 
\centering
\includegraphics[width=12cm,angle=-0]{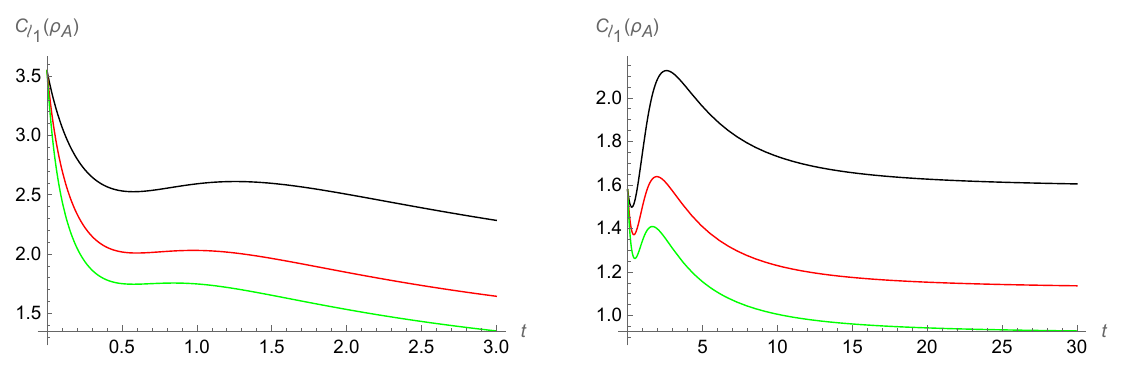}
\caption{
Evolution of the $\ell_1$-norm coherence of the reduced state $\rho_A$ for the case of common environment given by \eqref{eq: cl1-rhoA-com} for the given relaxation rate $\ga=0.1$ and different values of the squeezing parameter: $ s = 0 $ (left panel) and $ s = \log(10)/2$ (right panel). 
Color codes are as follows: $T=5$ (black), $T=10$ (red) and $T=15$ (green).
}
\label{fig: reducedcoh} 
\end{figure}

In figure \ref{fig: reducedcoh}, we present the time evolution of local $\ell_1$-norm coherence for the common environment scenario described in equation \eqref{eq: cl1-rhoA-com}. The left panel illustrates the non-squeezed state, while the right panel shows the squeezed state with a squeezing parameter of $ s = \log(10)/2 \approx 1.15 $, both for various temperature values.
In the case without squeezing, coherence starts at its maximum value at  $t=0$  and then begins to decline immediately. This decay is smooth and gradual, with the rate influenced by the temperature $T$ . Higher temperatures (green, red) result in a more rapid loss of coherence, while lower temperatures (black) allow coherence to persist for a longer duration. There exists a finite time interval during which a revival of coherence can be observed, with this effect being more pronounced at lower temperatures.
When squeezing is introduced, coherence initially decreases similarly to the scenario on the left. After reaching a minimum point, coherence begins to rise again for a limited time, indicating a revival effect. However, this increase is not permanent; after reaching a peak, coherence starts to decline once more toward its stationary value. A notable difference from the no-squeezing case emerges here: at lower temperatures (black and red), the peak coherence exceeds the initial value, demonstrating an induced coherence effect.
The minimum (maximum) occurs earlier (later) for lower temperatures confirming the longer-lasting and more pronounced revival before coherence vanishes while higher $T$ reduces the revival effect because thermal fluctuations dominate, making coherence loss more rapid.

\begin{figure} 
\centering
\includegraphics[width=12cm,angle=-0]{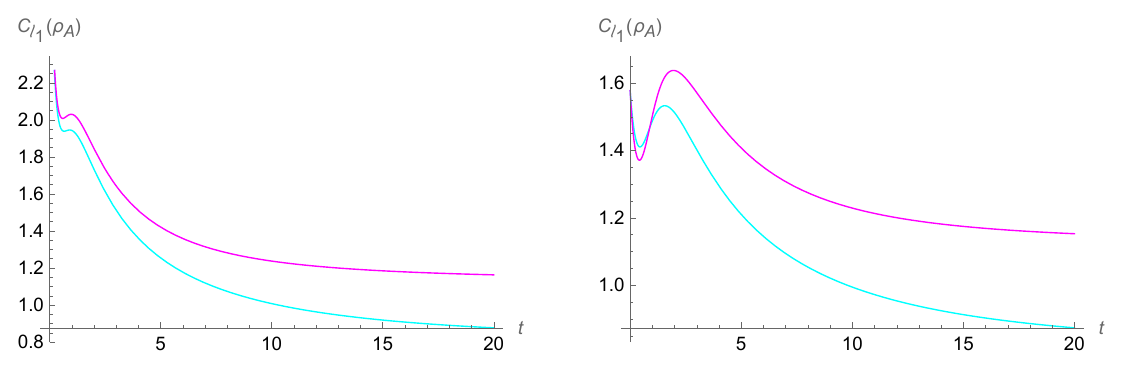}
\caption{
Evolution of the $\ell_1$-norm coherence of the reduced state $\rho_A$ for distinct environments (cyan curve) given by \eqref{eq: cl1-rhoA-dis} and for common environment (magenta curve) given by \eqref{eq: cl1-rhoA-com} for $\ga = 0.1$ and $T=10$ for $ s = 0 $ (left panel) and $ s = \log(10)/2 \approx 1.15 $ (right panel). 
}
\label{fig: reducedcohDisCom} 
\end{figure}

Figure \ref{fig: reducedcohDisCom} compares two types of environments-represented in cyan for distinct environments and magenta for a common environment scenario-at a specific temperature, highlighting the evolution of local coherence. The left panel depicts the non-squeezed state, while the right panel shows the squeezed state with a squeezing parameter of  $ s \approx 1.15 $.
%
In the non-squeezed state, coherence is consistently higher in common environments than in distinct ones. This is expected, as the original double CL equation includes an additional term that induces correlations between particles in the former case. 
The squeezing parameter $s$ modulates the transient behavior; there exists a brief time period during which coherence in the distinct scenario surpasses that of the common environment.
Looking at the analytical expressions \eqref{eq: cl1-rhoA-dis} and \eqref{eq: cl1-rhoA-com}, in the short-time regime, the coherence functions contain competing exponentials.
For distinct environments, time-dependent terms like $ e^{2\ga t} $ and $ e^{4\ga t} $ control the evolution while for the common environments, the expressions contain stronger terms like $ e^{8\ga t} $, which can lead to a more pronounced initial suppression of coherence.
Initially, the coherence in distinct environments decays more slowly than in the common environment. This is because the interaction terms in the common environment case include additional correlations, which initially enhance decoherence effects.
However, at later times, coherence in the common environment increases at a higher rate, leading to the eventual crossing point where the magenta curve, representing local coherence for the common environment, overtakes the cyan curve representing local coherence for the distinct environments.
For higher $s$, the transition region (where cyan is above magenta) may shift in time, but the long-term hierarchy remains unchanged. Since the stationary values are independent of $s$, the final dominance of the common environment case is not affected by squeezing.
Stationary coherence in a common environment is always $\sqrt{2}$ times larger than in distinct environments, highlighting an advantage in using common reservoirs for quantum coherence preservation.
%

\begin{figure} 
\centering
\includegraphics[width=12cm,angle=-0]{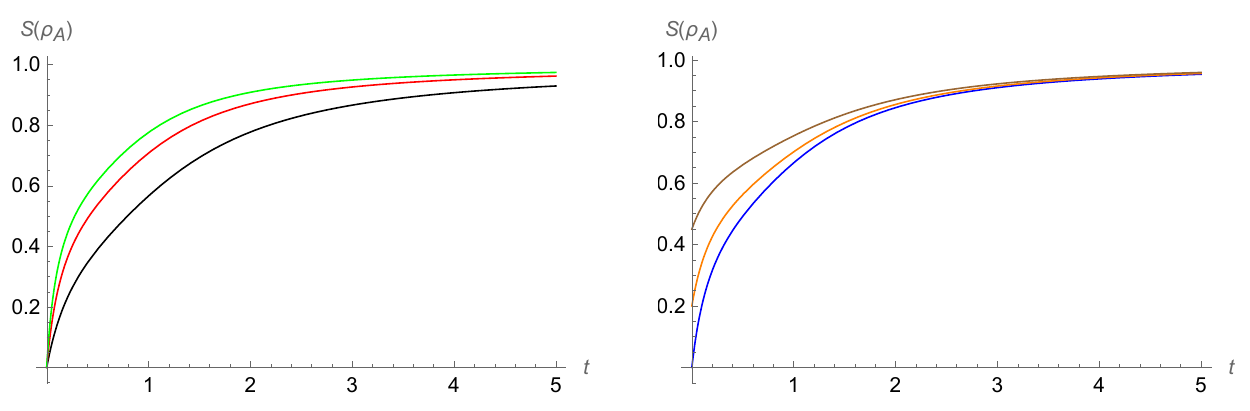}
\caption{
Evolution of the linear entropy of the reduced state $\rho_A$ for the case of common environment given by \eqref{eq: local_linent} and \eqref{eq: pure-com} for the given relaxation rate $\ga=0.1$.
Left panel considers different temperatures for the same squeezing parameter $ s = 0 $ while the right one considers different squeezing parameters for the same temperature $T=8$. 
Color codes are as follows: $T=5$ (black), $T=10$ (red) and $T=15$ (green), $s=0$ (blue), $ s = -\log(0.5)/2 \approx 0.35 $ (orange) and $ s = -\log(0.3)/2 \approx 0.6 $ (brown).
}
\label{fig: reducedentcom} 
\end{figure}

The plot shown in figure \ref{fig: reducedentcom} depicts the progression of the linear entropy $ S(\rho_A) $ for the reduced state in the context of a shared environment. The left panel examines various temperatures for the non-squeezed state ($s=0$), while the right panel focuses on different squeezed states, specifically the squeezed state described in Eq. \eqref{eq: squ0_pos}, with varying values of the squeezing parameter at the same temperature.
As shown in the left panel, the linear entropy increases more rapidly with rising temperature. At higher temperatures, the system undergoes stronger thermal fluctuations, resulting in quicker decoherence and a more mixed reduced state. In contrast, the right panel indicates that increasing squeezing leads to a slower rise in entropy. 
Higher levels of squeezing correspond to greater linear entropy, indicating that the local state becomes more mixed with increased squeezing. This suggests that squeezing intensifies the interaction between the system and its environment, thereby accelerating decoherence in the local state.

\begin{figure} 
\centering
\includegraphics[width=12cm,angle=-0]{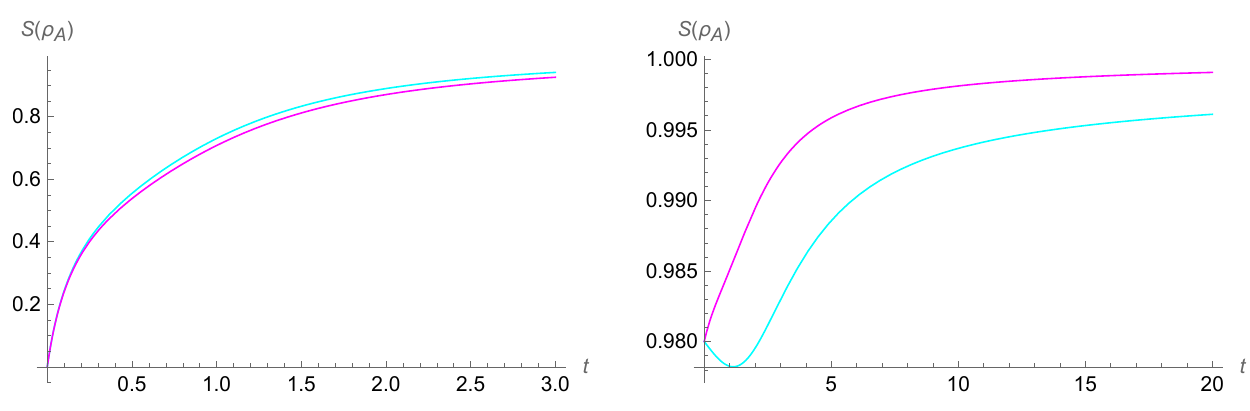}
\caption{
Evolution of the linear entropy of the reduced state $\rho_A$ for distinct environments (cyan curve) and for common environment (magenta curve) given by \eqref{eq: local_linent} and \eqref{eq: pure-dis}; and \eqref{eq: pure-com} for $\ga = 0.1$ and $T=10$ for $ s = 0 $ (left panel) and $ s = -\log(0.01)/2 \approx 2.3 $ (right panel). 
}
\label{fig: reducedentDisCom} 
\end{figure}

In Figure \ref{fig: reducedentDisCom}, we compare the two types of environmental scenarios in relation to the evolution of local linear entropy.
Unlike the left panel ($s=0$), where both curves increase monotonically, in the right panel ($s\approx 2.3$), the cyan curve (distinct environments) shows a short interval where it decreases before increasing again. 
One should not confuse this temporary entropy reduction with the same effect observed in a genuinely non-Markovian environment. In the latter case, entropy reduction results from actual memory effects, where lost information flows back from the bath, whereas in the former, it arises from initial entanglement in the squeezed state, which dynamically redistributes noise between the subsystems, momentarily suppressing local decoherence. As we discussed after Eq. \eqref{eq: linent-short-dis} this happens only when the condition $ T < \cosh(2s)/2 $ is satisfied.
%
In the common environment case (magenta curve), the subsystems are correlated through their shared environment. This strong correlation causes irreversible decoherence, preventing any temporary entropy reduction. As a result, the magenta curve remains consistently increasing, while the cyan curve experiences a brief decline.
It is important to point out that, as we will soon demonstrate in the case of a common environment, recoherence can be observed in the entanglement measured by log negativity.

\begin{figure} 
\centering
\includegraphics[width=12cm,angle=-0]{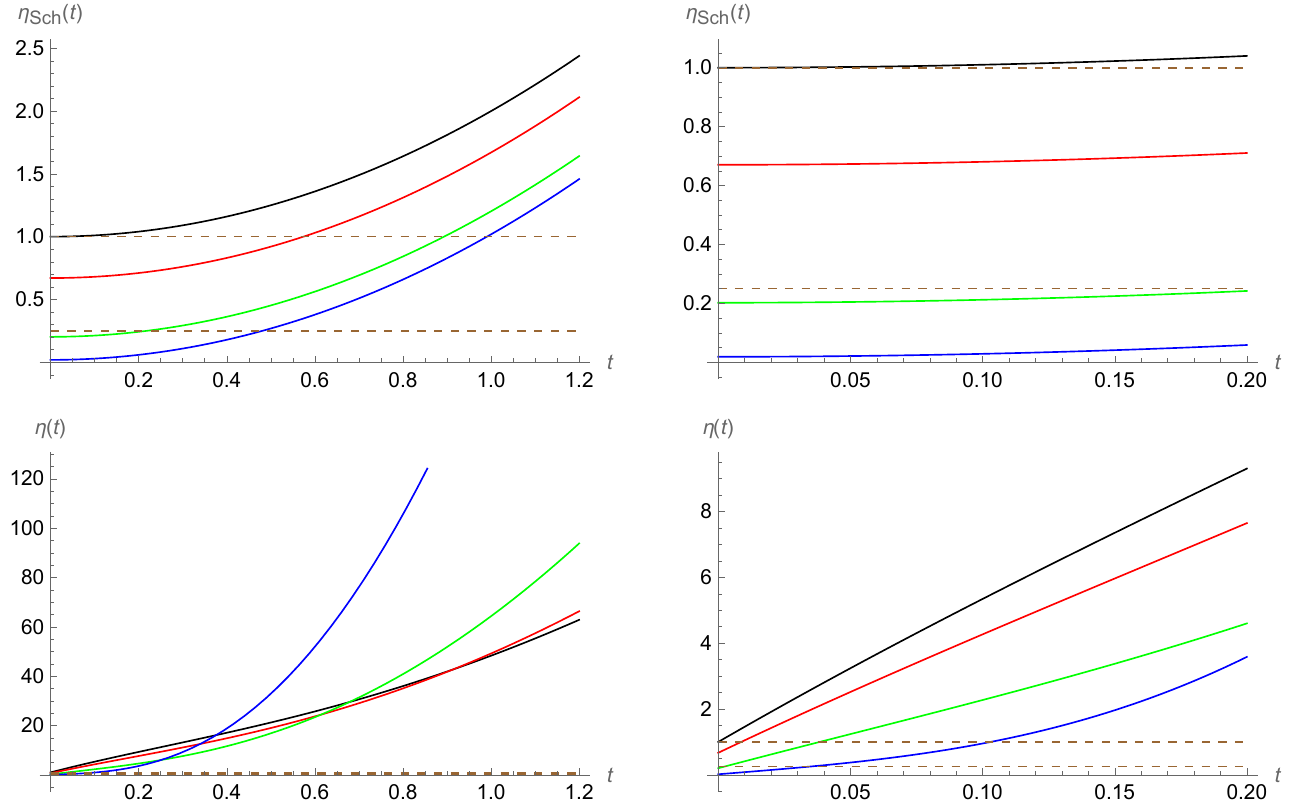}
\caption{
The progression of the EPR correlations $\eta(t)$ is analyzed within the context of the Schr\"odinger framework (top panels), as described by \eqref{eq: eta-Sch}, and within the framework of the CL with common environments (bottom panels), as outlined in \eqref{eq: eta-com}. Right panels are zoom-in views of the left ones for short times. This analysis pertains to a two-particle system initially represented by the squeezed state \eqref{eq: squ0_pos} for various values of the squeezing parameter: $ s=0 $ (black), $ s=0.1 $ (red), $ s=0.4 $ (green) and $ s=1 $ (blue). Parameters of the environment are $\ga =0.2 $ and $T=15$.
}
\label{fig: petaSchcom} 
\end{figure}

Figure \ref{fig: petaSchcom} represents the progression of the EPR correlation $\eta(t)$ within the context of the Schr\"odinger framework (right panel), as described by \eqref{eq: eta-Sch}, and within the framework of the CL with common environments, as outlined in \eqref{eq: eta-com}.
In the Schr\"odinger framework, as Eq. \eqref{eq: eta-Sch} shows $\eta(t)$ starts at $ \eta(0)=e^{-4s} $ meaning stronger squeezing ($s>0$) leads to smaller initial values, promoting entanglement.
The function grows quadratically with time due to the $t^2$ term, eventually exceeding the entanglement threshold ( $\eta=1$).
The right panel (zoomed-in) highlights the short-time behavior, where for $ s > \ln(4)/4 \approx 0.374 $, $\eta(0)<1/4$, indicating EPR correlations. However, this advantage diminishes over time as $\eta(t)$ increases.
In the CL context, $\eta(t)$ exhibits a more complex evolution due to dissipation and thermal effects in \eqref{eq: eta-com}. The presence of the environment introduces decoherence, which causes $\eta(t)$ to grow faster compared to the Schr\"odinger case.
Initially, squeezing still reduces $\eta(0)$, possibly achieving EPR correlations $\eta(0)<1/4$ for sufficiently large $s$.
However, the interaction with the environment leads to an increase in $\eta(t)$, degrading both entanglement and EPR correlations over time. The right panel shows that, at short times, squeezing can still enhance entanglement, but the influence of the environment becomes prominent at longer times.
%

\begin{figure} 
\centering
\includegraphics[width=12cm,angle=-0]{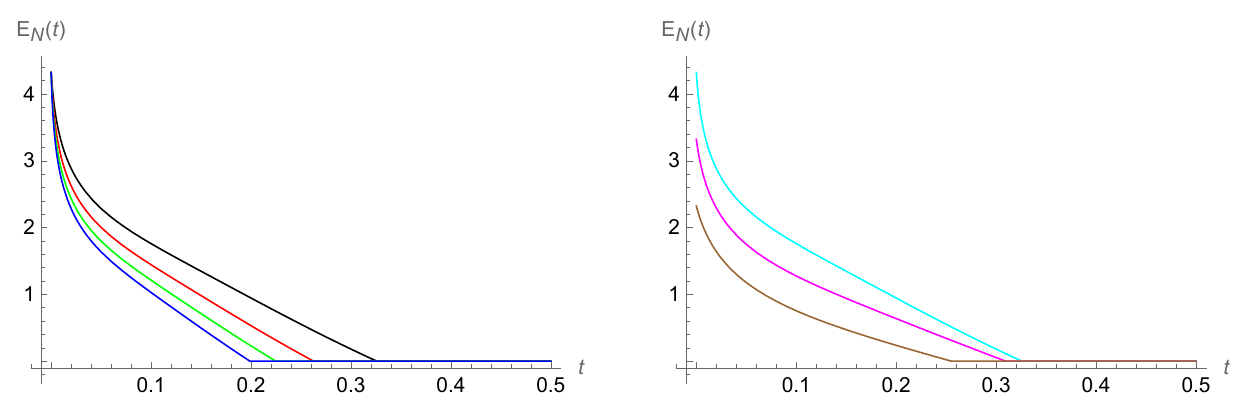}
\caption{
Evolution of the entanglement negativity \eqref{eq: log-neg} for distinct environments for $ \ga = 0.2 $ for $ s = \ln(20) / 2 \approx 1.5 $ (left panel) and for $T=10$ (right panel).
Color codes are as follows: $T=10$ (black), $T=15$ (red), $T=20$ (green) and $T=25$ (blue); $ s = \ln(20) / 2 \approx 1.5 $ (cyan), $ s = \ln(10) / 2 \approx 1.15 $ (magenta) and $ s = \ln(5) / 2 \approx 0.8 $ (brown).
}
\label{fig: pEntdis} 
\end{figure}

In Figure \ref{fig: pEntdis}, we have illustrated the evolution of entanglement log negativity as the system interacts with two distinct environments. 
The left panel illustrates the impact of varying temperatures on the entanglement dynamics for a specific squeezing parameter, while the right panel examines the influence of squeezing at a fixed temperature. 
In both panels, we observe a decay in entanglement negativity over time. There are clear points where the entanglement negativity drops to zero, indicating the phenomenon of entanglement sudden death. After reaching zero, entanglement does not revive, suggesting irreversible loss due to decoherence effects.
Higher temperatures accelerate entanglement decay, meaning that thermal noise destroys quantum correlations more quickly.
Lower temperatures retain entanglement for a longer time.
Higher squeezing results in stronger initial entanglement and a slower decay. Lower squeezing shows a faster decay, implying that initial squeezing helps sustain entanglement against decoherence.

\begin{figure} 
\centering
\includegraphics[width=12cm,angle=-0]{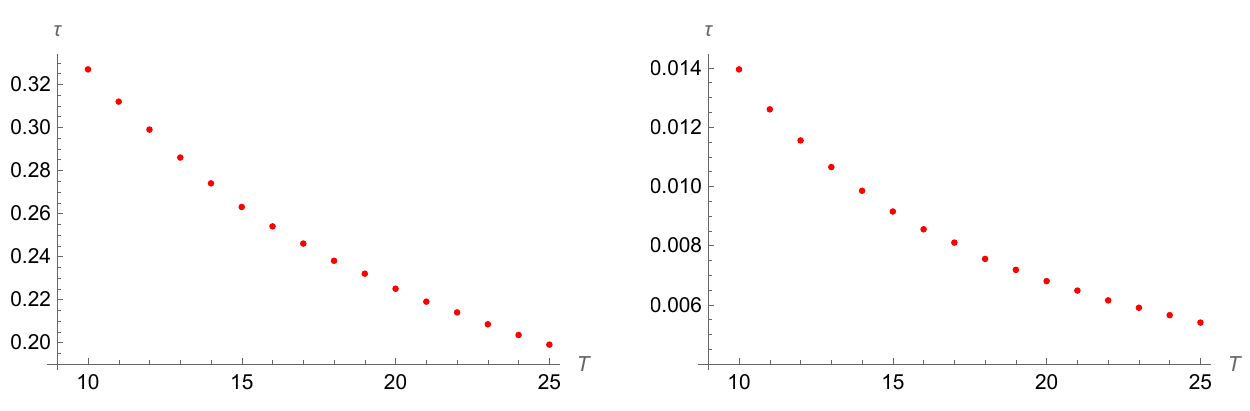}
\caption{
The time of entanglement sudden death for distinct environments for $\ga = 0.2$ in terms of temperature for $ s =  \ln(20) / 2 \approx 1.5  $ (left panel) and $ s =  -\ln(0.9) / 2 \approx 0.053  $ (right panel).
}
\label{fig: ESD} 
\end{figure}

In figure \ref{fig: ESD}, we present the entanglement death time as a function of temperature for two distinct values of the squeezing parameter, $ s \approx 1.5  $ (left panel) and $ s \approx 0.053  $ (right panel). Comparison of these panels show that higher squeezing delays entanglement sudden death, while lower squeezing results in almost immediate entanglement loss. Furthermore, temperature has a strong influence, but its effect is more pronounced in the weakly squeezed case. 

\begin{figure} 
\centering
\includegraphics[width=12cm,angle=-0]{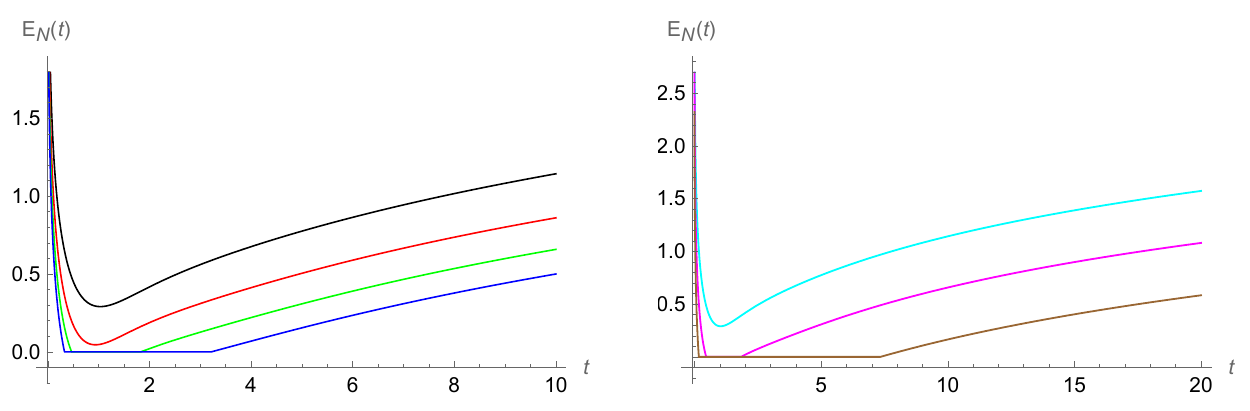}
\caption{
Evolution of the entanglement negativity \eqref{eq: log-neg} for common environments for $ \ga = 0.2 $ for $ s = \ln(20) / 2 \approx 1.5 $ (left panel) and for $T=10$ (right panel).
Color codes are as follows: $T=10$ (black), $T=15$ (red), $T=20$ (green) and $T=25$ (blue); $ s = \ln(20) / 2 \approx 1.5 $ (cyan), $ s = \ln(10) / 2 \approx 1.15 $ (magenta) and $ s = \ln(5) / 2 \approx 0.8 $ (brown).
}
\label{fig: pEntcom} 
\end{figure}

Figure \ref{fig: pEntcom} illustrates the log negativity of entanglement in the context of common environments. The figure shows that entanglement initially decreases for a period before increasing indefinitely due to the effective interaction between the particles induced by the common environment. Additionally, for specific values of the parameters related to the environment and the wavepacket, a dark period of entanglement is also observed.
From the left panel we find that the entanglement negativity remains significant over a longer time range compared to the distinct environments scenario. Higher temperatures still cause a faster decline in entanglement, but there is no abrupt entanglement sudden death as seen before. Furthermore, as before lower temperatures maintain entanglement longer.
Right panel shows that higher squeezing results in stronger and longer-lasting entanglement; and lower squeezing shows a faster decay, but the negativity does not immediately vanish.
Furthermore, specific parameter values associated with the environment and the wavepacket-namely temperature and squeezing-reveal a dark period of entanglement where entanglement vanishes. This behavior is a signature of non-Markovian effects arising from the common environment. The environment induces temporary loss of entanglement, but due to back-action or memory effects, entanglement re-emerges later. This is in stark contrast to distinct environments, where entanglement once lost is permanently destroyed.
Higher temperatures increase the likelihood of dark periods, suggesting that thermal fluctuations play a role in temporarily suppressing entanglement. 
In cases of higher squeezing, any dark period that do occur is shorter or may not happen at all, allowing entanglement to revive more rapidly. Conversely, with lower squeezing values, dark periods are longer. This suggests that stronger initial squeezing helps prevent or shorten dark periods, keeping entanglement alive.
Although not shown in the figures, calculations indicate that the non-squeezed state ($s=0$), which is initially disentangled, remains separable over time. This finding suggests that, in this specific case, the common environment does not generate entanglement, in contrast to squeezed states, where entanglement, after an initial decay or a dark period, can be enhanced.

In the Schr\"odinger framework, entanglement remains constant over time, which is expected since it must be invariant under local unitary transformations. When the particles in the system do not interact, the global unitary evolution is simply the tensor product of the local evolutions, expressed as $ \hat{U} = e^{- i (\hat{H}_1+\hat{H}_2)/\hb} = e^{- i \hat{H}_1/\hb} \otimes e^{- i \hat{H}_2/\hb} $. As a result, the entanglement negativity in this framework is given by
\begin{eqnarray}
E_N\big|_{\Sch} &=& \frac{2}{\ln(2)} s
\end{eqnarray}
which becomes zero only when the squeezing parameter $s$ is zero---corresponding to a separable state.

It is noteworthy that in quantum systems with discrete variables, a common bosonic bath has been utilized to generate entanglement between two qubits. The dual role of the environment---both generating entanglement and inducing decoherence---has been extensively studied \cite{OhKi-PRA-2006}.
In this context, the phenomenon of entanglement sudden death has been observed when each subsystem interacts with its own distinct environment \cite{YuEb-PRL-2004, YuEb-PRL-2006, EbYu-Sci-2007}. However, entanglement dynamics become more intricate in the presence of a common environment, where entanglement can revive after a finite dark period \cite{FiTa-PRA-2006}. For a recent investigation into the effects of environmental decoherence on various correlations in a two-qubit system, see \cite{Mo-EPJP-2024}.

\section{summary and conclusions} \label{sec: sum-con}

In this study, we investigated the dynamics of a free two-particle system initially prepared in a squeezed state, with no interaction between particles or with external agents. The system's evolution was modeled using the double Caldeira-Leggett equation in the high-temperature limit, treating the system as an open quantum system where environmental effects, such as dissipation and temperature, were incorporated. 
Notably, the presence of additional terms-absent in the von Neumann equation-leads to decoherence.
We explored two environmental coupling scenarios: (1) each particle interacts with a distinct environment and (2) both particles interact with a common environment. In the latter case, despite the presence of noise, the common environment generates correlations that contribute to maintaining or even amplifying entanglement. Our analysis considered identical damping coefficients and temperatures for distinct environments, and we also explored the Schr\"odinger limit, where damping and temperature are zero.
Key quantities studied include local quantum coherence, linear entropy, EPR correlations, and entanglement, the latter quantified via logarithmic negativity. We found that local coherence is directly proportional to the coherence length, defined as the width of the density matrix along the off-diagonal direction. Temperature accelerates decoherence, weakening the revival effect, while common environments preserve coherence more effectively than distinct ones. Squeezing plays a significant role in transient coherence dynamics but does not affect long-term saturation values.
Regarding entropy, high temperature and strong squeezing both promote a mixed local state, though their physical effects differ. High squeezing enhances initial correlations, which compete with environment-induced decoherence, resulting in transient reductions in entropy for systems with distinct environments.
EPR correlation  $\eta(t)$ showed that stronger squeezing improves initial entanglement in the Schr\"odinger framework, with $\eta(t)$ growing quadratically before surpassing the entanglement threshold. However, in the CL framework with common environments, decoherence accelerates the growth of $\eta(t)$, leading to a more rapid degradation of entanglement and EPR correlations over time. While squeezing enhances early-time correlations, the environment's influence becomes dominant at later times.
Entanglement behavior differed across the two scenarios. In systems with distinct environments, we observed entanglement sudden death, which occurred more rapidly at higher temperatures. In contrast, in a common environment, entanglement experiences a dark period at high temperatures and low squeezing, but this period shortens with stronger squeezing, suggesting that squeezing can delay or eliminate these dark periods, enabling entanglement to remain active or recover more rapidly.

In conclusion, this study provided a detailed comparison between systems interacting with distinct versus common environments, emphasizing how squeezing and decoherence influence key quantum properties. The results show that common environments are more effective at preserving coherence over time, while squeezing impacts the initial dynamics but does not affect long-term saturation. Future research could extend this analysis to more complex systems and explore strategies for optimizing environmental conditions to better preserve quantum correlations for potential quantum technologies.

\vspace{1cm}

{\bf{Data availability}}: This manuscript has no associated data.

{\bf{Conflict of Interest}}: The authors declare no conflict of interest.

{\bf{Acknowledgement}}:
We acknowledge Dr. Mokhtar Abbasi for fruitful discussions about the coherence of continuous variable states. Support from the University of Qom is acknowledged.



\end{document}